\documentstyle[12pt]{article}
\renewcommand
\baselinestretch{1.4}

\begin{document}

\hfill UB-ECM-PF 93/8

\hfill April 1993

\vspace*{3mm}

\begin{center}

{\LARGE \bf
Partition functions for the rigid string and
membrane at any temperature}
\vspace{4mm}

\renewcommand
\baselinestretch{0.8}

{\sc E. Elizalde}\footnote{E-mail address: eli @
ebubecm1.bitnet}, {\sc S. Leseduarte}, \\
{\it Department E.C.M., Faculty of Physics, University of
Barcelona, \\
Diagonal 647, 08028 Barcelona, Spain}, \\
 and  {\sc S. D. Odintsov}\footnote{On sabbatical leave from
Tomsk Pedagogical Institute, 634041 Tomsk, Russia. E-mail address:
odintsov @ theo.phys.sci.hiroshima-u.ac.jp} \\ {\it Department of
Physics, Faculty of Science, Hiroshima University, \\
Higashi-Hiroshima 724, Japan}
\medskip

\renewcommand
\baselinestretch{1.4}

\vspace{5mm}

{\bf Abstract}

\end{center}

Exact expressions for the partition functions of the
rigid string and
membrane
at any temperature are obtained in terms of hypergeometric
functions. By
using zeta function regularization methods, the results are
analytically
continued and written as asymptotic sums of Riemann-Hurwitz zeta
functions, which provide very good numerical approximations with
just a few first terms. This allows to obtain systematic
corrections to the results of Polchinski et al.,
corresponding to the limits
$T\rightarrow 0$ and $T\rightarrow \infty$ of the rigid string, and
to analyze the intermediate range of temperatures.
In particular, a way to obtain the Hagedorn temperature for
the rigid membrane  is thus found.

\vspace{8mm}


\newpage

\section{Introduction}

The problem of deciding if QCD is actually (some limit of) a string
theory (or, in general, of a theory of extended objects) is almost
twenty years
old and goes back to 't Hooft [1] (for recent reviews, see [2,3]).
In very new
 contributions, Polchinski and Yang have tried to answer this
question, both for the Nambu-Goto [4] and for the rigid string
cases
[5], by
calculating their partition functions and seeing if they actually
match with the partition function corresponding to QCD at different
ranges
of temperature. Due to the difficulties involved in the
calculations,
only a direct analysis of the two sharp limits
$T\rightarrow 0$ and $T\rightarrow \infty$ has been done. Such
approximations are rightfully
claimed to have overlapping validity. However, in the intermediate
region corrections to the first terms of the series expansion may
surely become important.

We shall here fill this gap, by obtaining exact analytical
continuations of
the partition functions, out of which approximations valid in
different
ranges of temperature will be quite easy to obtain. With
some additional
work, we will also be able to derive the corresponding expressions
for
bosonic membranes, both with and without rigid term. Extending the
ideas
of ref. [5] to the case of the rigid membrane, we can also
characterize
the Hagedorn transition in this case, and even for any kind of
membrane
and $p$-brane in general. As is known, this has always been a very
elusive subject.
The high-temperature limits of the free energy per unit area for
the
bosonic
membrane and for the rigid membrane are found not to agree with
that
of the QCD large-$N$ limit.

The paper is organized as follows. In sect. 2, after a very short
review of the results of Polchinski et al., we use our formalism
for the
calculation of an analytic partition function of
the rigid string which is exact to one-loop in the coupling
constant. In
two separate subsections, we obtain the low-temperature
and the high-temperature limits of the partition function,
respectively,
that is, the dominant terms plus first order corrections in each
case.
We also show explicitly how higher order terms can be obtained
---in a
way consistent with the loop expansion.
In sect. 3 we proceed with the calculation of the partition functions
corresponding to bosonic membranes and $p$-branes (first
subsection) and
to rigid membranes and rigid $p$-branes (second subsection).
The calculation of the extremum with respect to the parameters, in
each
case, is done in sect. 4. Establishing a parallelism with the
results of ref.
[5], we characterize the Hagedorn transition for the case of the
rigid
membrane and we study the very high temperature limits
corresponding to
the pure bosonic and  rigid membrane. Finally, sect. 5 is
devoted
to conclusions.

\bigskip

\section{Analytical partition function for the rigid string}

The first two terms in the loop expansion
\begin{equation}
S_{eff} = S_0+S_1+ \cdots
\end{equation}
 of the effective action corresponding to the rigid  string [6,7]
\begin{equation}
S= \frac{1}{2\alpha_0} \int d^2\sigma \, \left[ \rho^{-1}
\partial^2 X^\mu \partial^2 X_\mu + \lambda^{ab} \left( \partial_a
X^\mu \partial_b X_\mu - \rho \delta_{ab} \right) \right] + \mu_0
 \int d^2\sigma \, \rho,
\end{equation}
where $\alpha_0$ is the dimensionless, asymptotically free
coupling constant,
$\rho$ the intrinsic metric, $\mu_0$ the explicit string tension
(important at low energy) and $\lambda^{ab}$, $a,b=1,2$, the usual
Lagrange
multipliers, are given ---in the world sheet $0\leq \sigma^1 \leq
L$
and $0\leq \sigma^2 \leq \beta t$ (an annulus of modulus $t$)--- by
\begin{equation}
S_0= \frac{L\beta t}{2\alpha_0} \left[ \lambda^{11} +  \lambda^{22}
t^{-2} + \rho \left( 2\alpha_0 \mu_0 -  \lambda^{aa} \right)
\right]
\end{equation}
at tree level, and by
\begin{eqnarray}
S_1 &=& \frac{d-2}{2} \, \ln \det \left( \partial^4- \rho
\lambda^{ab} \partial_a  \partial_b \right) \nonumber \\
&=&  \frac{d-2}{2} \, L \sum_{n=-\infty}^{\infty} \int_{-
\infty}^{+\infty} \frac{dk}{2\pi} \, \ln \left[ \left( k^2 +
\frac{4\pi^2n^2}{\beta^2t^2} \right)^2 + \rho \left(  \lambda^{11}
k^2 + \frac{4\pi^2 \lambda^{22}n^2}{\beta^2t^2} \right)
\right] \end{eqnarray}
at one-loop order, respectively. Of course,  to make sense,
this
last expression
needs to be regularized and its calculation is  highly non-trivial
---as explicitly quoted in refs. [4,5]. There, it has been
obtained {\it only} in the very strict limits
$T\rightarrow 0$ and $T\rightarrow \infty$ (just terms of highest
order
have been kept in both cases) and around some extremizing
configuration of the parameters $\rho$, $\lambda^{11}$,
$\lambda^{22}$ and $t$. A semiclassical discussion of
rigid strings can be  found in refs. [8,9] also.

Here, we shall make use of the zeta function procedure [10], as
developed
in [11] after the rigorous proof of the zeta function
regularization
theorem, and further extended in [12] to the more elaborate
situations  we
need to deal with here. One can write
\begin{equation}
S_1= -(d-2) L \left. \frac{d}{ds} \zeta_A (s/2)\right|_{s=0}, \
\ \ \ \zeta_A (s/2) =   \sum_{n=-\infty}^{\infty} \int_{-
\infty}^{+\infty} \frac{dk}{2\pi} \, (k^2+y_+^2)^{-s/2}  (k^2+
y_-^2)^{-s/2},
\end{equation}
where
\begin{equation}
y_{\pm} = \frac{a}{t} \left[ n^2 + \frac{\rho
t^2\lambda^{11}}{2a^2} \pm \sqrt{\rho} \, \frac{t}{a} \left(
(\lambda^{11}-\lambda^{22}) n^2 + \frac{\rho
t^2\lambda^{11^2}}{4a^2} \right)^{1/2} \right]^{1/2}, \ \ \ a\equiv
\frac{2\pi}{\beta}.
\end{equation}
Following refs. [4,5], we may consider two basic approximations
of overlapping
validity: one for low temperature, $\beta^{-2} << \mu_0$, and
the other for high temperature,  $\beta^{-2} >> \alpha_0\mu_0$.
Both these approximations (overlapping region included) can be
obtained
from the expression above, which on its turn can be written in the
form
\begin{equation}
 \zeta_A (s/2) = \frac{1}{2\sqrt{\pi}} \, \frac{\Gamma (s-
1/2)}{\Gamma (s)}\,  \sum_{n=-\infty}^{\infty} \frac{y_-
}{(y_+y_-)^s} \, F(s/2,1/2;s; 1-\eta), \ \ \ \ \eta\equiv \frac{y_-
^2}{y_+^2}.
\label{zz1}
\end{equation}
This is an {\it exact} formula. To finish this short introduction,
we
should point out that the  above approach is somewhat
different from the standard approach to the free energy of the
bosonic
string at non-zero temperature (for a review and a list of
references see, for example, [14]).
\medskip

\subsection{The low temperature approximation}

The low temperature case is already quite involved and the refined
methods of ref. [12] must be used. The term $n=0$ in (\ref{zz1}) is
non-vanishing and must be treated separately from the rest. It
gives
\begin{equation}
\zeta_A^{n=0} (s/2) = \frac{1}{2\pi} \, \frac{\Gamma ((1-s)/2)
\Gamma(s-1/2)}{\Gamma (s/2)} \, (\lambda^{11} \rho )^{1/2-s}, \ \
\ \ \frac{1}{2} < \Re (s) < 1.
\end{equation}
This is again an exact expression, that yields
\begin{equation}
 \left. \frac{d}{ds} \zeta_A^{n=0} (s/2)\right|_{s=0} =-
\frac{1}{2} \, \sqrt{\rho \lambda^{11}}
\end{equation}
and
\begin{equation}
S_1^{(n=0)} =  \frac{d-2}{2} \, \sqrt{\rho \lambda^{11}}.
\label{s10}
\end{equation}
This contribution must be added to the one coming from the
remaining terms in eq. (\ref{zz1}) above
\begin{equation}
 \zeta_A' (s/2) = \frac{1}{\sqrt{\pi}} \, \frac{\Gamma (s-
1/2)}{\Gamma (s)}\,  \sum_{n=1}^{\infty} \frac{y_-
}{(y_+y_-)^s} \, F(s/2,1/2;s; 1-\eta), \ \ \ \
\Re (s) >1
\label{zp1}
\end{equation}
(the prime is here no derivative, it just means that the term $n=0$
is
absent from the sum). Within this approximation, and working around
the
classical, $T=0$ solution: \\   $\lambda^{11}$ $ =\lambda^{22} $ $
=\alpha_0 \mu_0$, $\rho = t^{-2} =1$, we obtain
\begin{equation}
 \zeta_A' (s/2) = \frac{1}{\sqrt{\pi}} \, \frac{\Gamma (s-
1/2)}{\Gamma (s)}\, \left(\frac{a}{t} \right)^{1-2s} \left[ F_0
(s)+F_1 (s)+F_2 (s)  \right],
\label{zp2}
\end{equation}
where
\begin{eqnarray}
&& F_0 (s) \equiv  \sum_{n=1}^{\infty} \frac{n^{1-
s}}{(n^2+\sigma_2^2)^{s/2}} \,
\left[F(s/2,1/2;s;\sigma_1^2/(n^2+\sigma_1^2)) -1 -
\frac{\sigma_1^2}{4n^2} \right],  \\
&& F_1 (s) \equiv  \sum_{n=1}^{\infty} \frac{n^{1-
s}}{(n^2+\sigma_2^2)^{s/2}}, \ \ \ \ F_2 (s) \equiv
\frac{\sigma_1^2}{4} \, \sum_{n=1}^{\infty} \frac{n^{-1-s
}}{(n^2+\sigma_2^2)^{s/2}}, \ \ \ \  \sigma_i^2\equiv
\frac{\lambda^{ii}
\rho t^2\beta^2}{4\pi^2}, \ i=1,2. \nonumber
\label{efes}
\end{eqnarray}
In this formula we have already taken advantage of the low energy
limit
and from now on in this subsection we shall compute perturbations
around it.

The study of the functions (13) is done in [12] in full
detail. Substituting the results back into  (\ref{zp2}), we
obtain [12]
\begin{eqnarray}
 \zeta_A' (s/2) & =&  \frac{1}{\sqrt{\pi}} \, \Gamma (s-
1/2)\, \left(\frac{a}{t} \right)^{1-2s} \left[ 1+ \gamma s + {\cal
O} (s^2)\right] \nonumber \\  && \times \, \left[ sF_0(0)+
\frac{\sigma_1^2}{8}+ \frac{\sigma_1^2 \gamma s}{4}
-\frac{\sigma_2^2s}{4} - \frac{s}{12}
+ {\cal O} (s^2)\right],
\end{eqnarray}
therefore
\begin{equation}
 \zeta_A' (0) = -\frac{ a\sigma_1^2}{4t}
\end{equation}
and
\begin{eqnarray}
 \left. \frac{d}{ds} \zeta_A' (s/2)\right|_{s=0} &=& - \frac{a}{t}
\left[ \psi (-1/2) \frac{\sigma_1^2}{4}-  \ln (a/t)
\frac{\sigma_1^2}{2} + \frac{(1+\gamma)\sigma_1^2}{4} -
\frac{\sigma_1^2}{2} \ln
(\sigma_1 /2) -\frac{\sigma_1}{2} \right. \nonumber \\ &-& \left.
\frac{\sigma_2^2}{2}  - \frac{1}{12} +  \frac{1}{\pi}
\int_{\sigma_1}^{\infty} dr \, \ln
\left( 1- e^{-2\pi r} \right) \frac{r}{\sqrt{r^2-\sigma_1^2}}
\right].
\end{eqnarray}
Finally
\begin{eqnarray}
S_1^{(n\neq 0)}  &=& (d-2) \frac{La}{4t} \left\{ \sigma_1^2
\left[ - 2\ln (a \sigma_1 /t)+ 3
\right] - 2\sigma_1 \right. \nonumber \\ &-& \left. 2\sigma_2^2 -
\frac{1}{3} +
\frac{4}{\pi}  \int_{\sigma_1}^{\infty} dr \, \ln \left( 1-
e^{-2\pi r} \right) \frac{r}{\sqrt{r^2-\sigma_1^2}} \right\}.
\label{s1n0}
\end{eqnarray}

Adding now (\ref{s10}) and (\ref{s1n0}) we get the desired
expansion  of the one loop effective action near $T=0$. It reads
\begin{eqnarray}
S_1&=& \frac{(d-2)L}{2} \sqrt{\rho\lambda^{11} } + \frac{(d-2)\pi
L}{\beta t} \left[ -\sigma_1^2 \ln \left(
\frac{\sqrt{\rho\lambda^{11}}}{\mu } \right) +
\frac{3\sigma_1^2}{2} \right. \nonumber \\  &-& \left.
 \sigma_1 -\sigma_2^2 -
\frac{1}{6} +
\frac{2}{\pi}  \int_{\sigma_1}^{\infty} dr \, \ln \left( 1-
e^{-2\pi r} \right) \frac{r}{\sqrt{r^2-\sigma_1^2}} \right],
\end{eqnarray}
where $\mu$ is the ordinary mass constant that appears in the
general procedure of zeta-function regularization [13].

We can now obtain the extremum of $S_0+S_1$:
\begin{eqnarray}
\rho &\simeq & 1+  \frac{(d-2) \sqrt{\alpha_0}}{2\beta
\sqrt{\mu_0}}, \nonumber \\
\frac{1}{t^2} &\simeq & 1+  \frac{(d-2) \sqrt{\alpha_0}}{2\beta
\sqrt{\mu_0}}+  \frac{(d-2)\alpha_0 }{4\pi} \, \left[
\ln \left( \frac{\alpha_0 \mu_0}{\mu } \right) -3
 \right], \nonumber \\
\lambda^{11} &\simeq & \alpha_0 \mu_0 + \frac{(d-
2)\alpha_0^{3/2} \sqrt{\mu_0}}{4\beta}
, \\
\lambda^{22} &\simeq & \alpha_0 \mu_0 - \frac{(d-
2)\alpha_0^{3/2} \sqrt{\mu_0}}{4\beta}
. \nonumber
\end{eqnarray}
These are the first order corrections to the classical result,
 which corresponds to the $T \rightarrow 0$ case. In a calculation
 to two loop order it would have sense to go further with the
calculation of second order corrections. As is clear from the above
expressions, this can be done easily.
\medskip

\subsection{The high temperature approximation}

For high temperature, the ordinary expansion of the confluent
hypergeometric function $F$ of eq. (\ref{zz1}) is in order
\begin{equation}
F(s/2,1/2;s; 1-\eta) = \sum_{k=0}^{\infty}
\frac{(s/2)_k(1/2)_k}{k!(s)_k} \, (1-\eta )^k,
\end{equation}
$(s)_k = s(s+1) \cdots (s+k-1)$ being Pochhamer's symbol (the
rising factorial).
This series expansion is here
completely consistent with the loop approximation ($\alpha_0$ is
kept always small) and, actually, only the first
three terms of this expansion need to be taken into account
at the approximation in which we are working. However, the fact
that there is no mathematical hindrance against going further with
these formulas must be properly remarked.
It is useful for the future calculation to notice that
\begin{equation}
\lim_{s\rightarrow 0}  F(s/2,1/2;s; 1-\eta) =  1 + \eta^{-
1/2}.
\end{equation}
We can approximate $y_\pm$ by
\begin{equation}
y_\pm \simeq  \frac{a}{t} \,  \sum_{n=-\infty }^{\infty} \left[ (n
\pm b)^2+c^2 \right]^{1/2}, \ \   b=\frac{\beta t}{4\pi}
\sqrt{\rho
(\lambda^{11}- \lambda^{22} )}, \ \ c^2= \frac{\rho \beta^2 t^2
(\lambda^{11}+ \lambda^{22} )}{16\pi^2} \simeq
\frac{\alpha_0^2}{16\pi^2}.
\label{yes}
\end{equation}
Before taking the derivative of the zeta function (\ref{zz1})
one should
observe that, apart from the usual contribution coming from the
derivative
of $\Gamma (s)^{-1}$ at $s=0$, there will be here an additional
term
due
to the pole of the Hurwitz zeta function $\zeta (1, \pm b)$ which
emerges from  the $y_\pm$ in (\ref{zz1}).
After  appropriate analytic continuation, the derivative of the
zeta function yields
\begin{equation}
 \left. \frac{d}{ds} \zeta_A (s/2)\right|_{s=0}
= -\frac{1}{4} \sqrt{b^2+c^2} +\frac{2\pi }{\beta t} \left\{ b^2+
\frac{1}{6} - \left[ \frac{1}{2} \ln (-b^2) + 2-2\ln 2
 \right] c^2 \right\}.
\end{equation}
In order to obtain this result, which comes through elementary
Hurwitz
zeta functions, $\zeta (\pm 1,\pm b)$, we have used the binomial
expansion in (\ref{yes}). This is
completely consistent with the approximation employed (notice, once
more,
the extra terms coming from the contribution of the pole in
(\ref{yes})
to the derivative of $\zeta_A (s/2)$).

 It is immediate to see that the terms of highest order coincide
with the result of Polchinski and Yang
\begin{equation}
S_1= -\frac{d-2}{2} L \left[ \frac{2\pi}{3\beta t} + \sqrt{\rho
\lambda^{11}} + {\cal O} (\beta) \right].
\end{equation}
The outcome of the extremization of $S_0+S_1$ with respect to the
parameters  $\rho$, $\lambda^{11}$, $\lambda^{22}$ and $t$ is
\begin{eqnarray}
& & \rho^{-1} =t^2= 1-\frac{(d-2) \alpha_0}{2\beta} \,
\sqrt{\lambda^{11}}, \nonumber \\
& & \lambda^{22}= -\frac{(d-2) \alpha_0}{4\beta} \,
\sqrt{\lambda^{11}}+ \alpha_0\mu_0 +
\frac{(d-2)\pi \alpha_0}{3\beta^2},  \\
& & \lambda^{11}= \frac{3(d-2) \alpha_0}{4\beta} \,
\sqrt{\lambda^{11}}+ \alpha_0\mu_0 -
\frac{(d-2)\pi \alpha_0}{3\beta^2},  \nonumber
\end{eqnarray}
where it is this last equation the one which determines the
possible values of
$\lambda^{11}$. Thus, the values of the parameters at the two
transition points that appear at high temperature ---namely
\begin{equation}
\beta_c^2 \mu_0 = \frac{(d-2)\pi}{3}- \frac{(d-2)^2\alpha_0}{8},
\end{equation}
corresponding to the Hagedorn transition, and
\begin{equation}
\beta_{c'}^2 \mu_0 = \frac{(d-2)\pi}{3}- \frac{9(d-
2)^2\alpha_0}{64},
\end{equation}
beyond which the variables acquire an imaginary part--- and the
value of the winding-soliton mass squared
\begin{equation}
M^2 (\beta)^2 = \beta^2 t^2 (\lambda^{11})^2 \alpha_0^{-2},
\end{equation}
are only modified by higher order terms in
$\alpha_0$ (that is, of order $\sqrt{\alpha_0}$ in the last case).
In fact, the result for $S_1$ to next order is
\begin{eqnarray}
S_1^{(2)}&=& (d-2)L \left\{ - \frac{\pi}{3\beta t} +\frac{1}{2}
\sqrt{ \rho \lambda^{11}}- \frac{1}{8\pi}\, \beta t \rho
(\lambda^{11}- \lambda^{22} )  \right. \nonumber \\ &+& \left.
\frac{1}{8\pi}\, \beta t \rho (\lambda^{11}+ \lambda^{22} )
 \left[ \frac{1}{2} \ln \left( -\frac{\beta^2 t^2 \rho
(\lambda^{11}- \lambda^{22} )}{16\pi^2\mu} \right) + 2-2\ln 2
  \right] \right\}.
\end{eqnarray}
{}From the expression of the total action, $S$, to this order, new
equations for the extremizing configuration follow. In particular,
from
the derivatives of $S$ with respect to $\lambda^{22}$ and
$\lambda^{11}$ we obtain, respectively
\begin{eqnarray}
\frac{1}{t^2} & \simeq & \rho - \frac{d-2}{4\pi} \, \alpha_0 \rho
\left[  \frac{1}{2} \ln \left( \frac{(d-2) \alpha_0}{24\pi^2\mu}
\right) + \psi (-1/2) + \gamma +1 - \frac{3}{8} \beta
\sqrt{\lambda^{11}} \right], \nonumber \\
\frac{1}{\rho} & \simeq & 1-  \frac{(d-2) \alpha_0}{ \beta
\sqrt{\lambda^{11}}}-  \frac{(d-2) \alpha_0}{4\pi} \left[
\frac{1}{2} \ln \left( \frac{(d-2) \alpha_0}{24\pi^2\mu} \right) +
1- 2\ln 2 \right. \nonumber \\
&-& \left.  \frac{3}{8} \beta \sqrt{\lambda^{11}} + \frac{(d-2)
\alpha_0}{ \beta \sqrt{\lambda^{11}}} \left[ 3- 2\ln 2
 \right] \right].
\end{eqnarray}
One can perform a consistent evaluation in terms of the dominant
contribution (lowest power in $\alpha_0$) obtained in ref. [5]
\begin{eqnarray}
\frac{1}{t^2} & = & 1 +  \frac{(d-2) \alpha_0}{2 \beta
\sqrt{\lambda^{11}}}-  \frac{(d-2) \alpha_0}{2\pi} + {\cal O}
(\alpha_0^{3/2}, \alpha_0^{1/2} \mu_0 \beta^2), \nonumber \\
\frac{1}{\rho} & = & 1 -  \frac{(d-2) \alpha_0}{2 \beta
\sqrt{\lambda^{11}}}-  \frac{(d-2) \alpha_0}{4\pi} \left[
\frac{1}{2} \ln \left( \frac{(d-2) \alpha_0}{24\pi^2\mu} \right) +
1- 2\ln 2 \right] \\
& +& {\cal O} (\alpha_0^{3/2}, \alpha_0^{1/2} \mu_0 \beta^2).
\nonumber
\end{eqnarray}
On the other hand, from the derivatives with respect to $\rho$ and
$t$
(and within the same approximation) we get, respectively
\begin{eqnarray}
\lambda^{22} & = & 2\alpha_0 \mu_0 -\lambda^{11} +
\frac{d-2}{2\beta} \, \alpha_0 \sqrt{ \lambda^{11}} +
\frac{(d-2)^2 \alpha_0^2}{6\pi\beta^2} \left( 1- \frac{3}{2} \,
\beta \sqrt{ \lambda^{11}} \right) \nonumber \\
&+& \frac{(d-2)^2 \alpha_0^2}{4\pi\beta^2} \left( \beta \sqrt{
\lambda^{11}}+ \frac{\mu_0\beta^2}{d-2} \right)  \left[
\frac{1}{2} \ln \left( \frac{(d-2) \alpha_0}{24\pi^2\mu} \right) +
\frac{3}{2}- 2\ln 2 \right], \nonumber \\
\lambda^{11} & = & -\frac{(d-2)\pi \alpha_0}{3\beta^2} +\alpha_0
\mu_0 +\frac{3(d-2) \alpha_0}{4\beta}  \sqrt{ \lambda^{11}} \\
&-& \frac{(d-2) \alpha_0^2\mu_0}{4\beta  \sqrt{ \lambda^{11}}} -
\frac{3(d-2)^2\pi \alpha_0^2}{24\beta^3  \sqrt{ \lambda^{11}}}
 + {\cal O} (\alpha_0^2, \alpha_0 \mu_0 \beta^2).
\nonumber
\end{eqnarray}
Of course, the terms of order $\alpha_0^2$ and  $\alpha_0 \mu_0
\beta^2$ are not to be taken seriously, because they are of the
same order as the first ones that would come from the two-loop
contribution to the total action.

More than the actual expressions themselves, the thing to
be remarked here is the fact that we have constructed a simple
procedure to obtain
all higher-order contributions to the loop action and,
correspondingly,
to the extremizing configuration, the Hagedorn temperature, etc. in
a systematic,
rigorous, analytical way, both for low and for high temperature
and, taking the necessary terms, for any temperature in the
intermediate, overlapping region which connects both limits.
In the next section we shall extend our method to membranes and
$p$-branes.
 \bigskip

\section{Partition function for the bosonic membrane and p-brane}

We shall consider the cases of the pure bosonic membrane and of the
bosonic membrane plus rigid term, and also of their generalizations
in the form of $p$-branes.
The classical membrane theory has been formulated in refs. [15].
Semiclassical approaches to membrane theory can be found in refs.
[16-19], and general reviews on membranes are [18,23]. One should
also notice that string theories can be obtained as some
compactification of  membrane theories [24].

\subsection{Pure bosonic membrane and corresponding p-brane}

The tree level action similar to (3) is
\begin{equation}
S_0^{(m)} = \kappa L^2 \beta t \left[ (1+\sigma_0)^{1/2}
(1+\sigma_1) -
 \left( \frac{1}{2} \lambda_0 \sigma_0 +  \lambda_1 \sigma_1
\right) \right]
\label{tlm}
\end{equation}
where $\lambda_0$ and $\lambda_1$ are Lagrange multipliers and
$\sigma_0$ and $\sigma_1$ are composite fields, defined in [16,17].
The one-loop contribution to the action can be written formally as
follows (cf. [16,17])
 \begin{equation}
S_1^{(bm)}= \frac{(d-3)L^2}{2}  \sum_{n=-\infty}^{\infty} \int_{-
\infty}^{+\infty} \frac{dk_1dk_2}{(2\pi)^2}  \left[
\lambda_1 (k_1^2  +k_2^2  ) +
\frac{4\pi^2\lambda_0 n^2}{\beta^2t^2}  \right].
\end{equation}
This expression must be regularized. As in the string case, we
choose the zeta function method. Calling $\zeta_2$ the
corresponding zeta function, we have
\begin{eqnarray}
S_1^{(bm)} &=& - \frac{(d-3)L^2}{2} \left. \frac{d}{ds} \zeta_2 (s)
\right|_{s=0}, \nonumber \\
\zeta_2 (s) &=&  \sum_{n=-\infty}^{\infty} \int_{-
\infty}^{+\infty} \frac{dk_1dk_2}{(2\pi)^2} \, \ln \left[
\lambda_1 (k_1^2  +k_2^2  ) +
\frac{4\pi^2\lambda_0 n^2}{\beta^2t^2}  \right]^{-s}.
\end{eqnarray}
After some calculations, we get
\begin{equation}
\zeta_2 (s) = \frac{1}{4 \pi (s-1) \lambda_1} \, \left(
\frac{4\pi^2 \lambda_0}{\beta^2t^2} \right)^{1-s} \, \zeta_R (2s-
2),
\end{equation}
where $\zeta_R$ is Riemann's zeta function. We thus obtain
\begin{equation}
S_1^{(bm)} = - \frac{2 (d-3) \pi \lambda_0 L^2}{\lambda_1
\beta^2t^2} \,
\zeta_R' (-2).
\end{equation}
A  review of membrane theory at non-zero temperature
can be found in ref. [21].

In the case of the bosonic $p$-brane, the corresponding expressions
are
\begin{equation}
S_0^{(p)}= \kappa L^p \beta t \left[ (1+\sigma_0)^{1/2}
(1+\sigma_1)^{p/2} - \frac{1}{2} \left( \lambda_0 \sigma_0 + p
\lambda_1 \sigma_1 \right) \right]
\end{equation}
and
\begin{eqnarray}
S_1^{(bp)}&=&-\frac{(d-p-1)L^p}{2} \left. \frac{d}{ds} \zeta_p (s)
\right|_{s=0}, \nonumber \\
\zeta_p (s) &=&  \sum_{n=-\infty}^{\infty} \int_{-
\infty}^{+\infty} \frac{dk_1\cdots dk_p}{(2\pi)^p} \, \ln \left[
 \lambda_1 (k_1^2+ \cdots  +k_p^2  ) +
\frac{4\pi^2\lambda_0 n^2}{\beta^2t^2}  \right]^{-s} \nonumber \\
&=& \frac{V_p \Gamma (p/2) \Gamma (s-p/2)}{(2\pi )^p
\lambda_1^{p/2} \Gamma (s)}  \left( \frac{4\pi^2 \lambda_0
}{\beta^2t^2} \right)^{p/2-s} \, \zeta_R (2s-p),
\end{eqnarray}
where $V_p$ is the `volume' of the $p-1$-dimensional unit sphere.
Observe that the membrane is the particular case $p=2$ of this
formula (as it should). We get
\begin{equation}
S_1^{(bp)}= - \frac{d-p-1}{2} \frac{V_pL^p \Gamma(p/2) }{\beta^p
t^p}
\left( \frac{\lambda_0}{\lambda_1} \right)^{p/2} \left\{ \begin{array}{ll}
\displaystyle \frac{(-1)^{p/2}}{(p/2)!} \, \zeta_R' (-p), & p = \mbox{even},
\\
\Gamma(-p/2) \zeta_R (-p), & p = \mbox{odd}. \end{array} \right.
\end{equation}
We ought to remark here that very nice expressions for the
derivatives
of the Riemann and Hurwitz zeta functions at any negative integer
value of the argument have been obtained in [22].
\medskip

\subsection{Bosonic membrane with rigid term and the corresponding
p-brane}

The tree level action is the same as before, eq. (\ref{tlm}). The
one-loop order contribution for the bosonic membrane is
(for a semiclassical approach to the rigid $p$-brane see [20])
\begin{eqnarray}
S_1^{(rm)}&=& - \frac{(d-3)L^2}{2} \left. \frac{d}{ds} \zeta_2^r
(s)
\right|_{s=0},  \\
\zeta_2^r (s) &=&  \sum_{n=-\infty}^{\infty} \int_{-
\infty}^{+\infty} \frac{dk_1dk_2}{(2\pi)^2}   \left[
\frac{1}{\rho^2} \left(
k_1^2  +k_2^2 +
\frac{4\pi^2 n^2}{\beta^2t^2}\right)^2 + \kappa \left( \lambda_1
(k_1^2  +k_2^2  ) +
\frac{4\pi^2\lambda_0 n^2}{\beta^2t^2} \right)  \right]^{-s}, \nonumber
\end{eqnarray}
where the label $r$ means {\it rigid}.
We can write
\begin{equation}
\zeta_2^r (s)= \frac{\rho^{2s}}{4\pi}  \sum_{n=-\infty}^{\infty}
\int_0^{\infty} dk \, (k+c_+)^{-s} (k+c_-)^{-s},
\end{equation}
with
\begin{equation}
c_{\pm} = \frac{4\pi^2 n^2}{\beta^2t^2} + \frac{\kappa
\rho^2\lambda_1}{2} \pm \sqrt{\kappa}\, \rho \left[
(\lambda_1-\lambda_0)
 \frac{4\pi^2 n^2}{\beta^2t^2} + \frac{\kappa \rho^2\lambda^2_1}{4}
\right]^{1/2}.
\label{cpm}
\end{equation}
Here, in analogy with the rigid string case, the term
corresponding
to $n=0$ must be treated separately. It yields a beta function.
Also
as in the rigid string case, the remaining series can be written
in terms of a confluent hypergeometric function. The
complete result, obtained after some work, is
\begin{equation}
\zeta_2^r (s)= \frac{\rho^{2-2s}(\kappa \lambda_1)^{1-2s} \Gamma (1-s)
\Gamma (2s-1)}{4\pi \Gamma (s)} +  \frac{\rho^{2s} \Gamma (2s-
1)}{2\pi \Gamma (2s)} \sum_{n=1}^\infty c_-^{1-2s} F(s,2s-1;2s;1-
c_+/c_-).
\end{equation}

Again, it is remarkable that one can in fact obtain the {\it exact}
expressions corresponding to bosonic membranes and $p$-branes, both
for the ordinary and for the rigid case. The last one is really
intricate, specially for dealing with membranes and
 $p$-branes with $p$ even. In fact, when taking the derivative of
the zeta function at $s=0$ two alternative procedures can be
followed. This stems from the observation that, in the end, we
shall be interested in the high temperature approximation. Thus, we
can either start by making the corresponding expansion in the
confluent hypergeometric function, do then the analytical
continuation and finally take the derivative, or else, we can keep
the hypergeometric function exact during the process of analytical
continuation, then take the derivative, and make the
high-temperature expansion at the end. Both procedures have been
seen to yield exactly the same result. The second one is much more
involved
but has the advantage of keeping exact expression till the end. It
can be considered as zeta function regularization {\it at its
best}. No other procedure can compete with it in rigor and
elegance. However, let us repeat that the price to be paid is not
worth in many cases ---as the present, where for the
high-temperature analysis in which we are interested the first
alternative is much easier. We shall follow it here.

For the general case of the $p$-brane with rigid term, the one-loop
contribution to the action is
\begin{eqnarray}
S_1^{(rp)} &=& - \frac{(d-p-1)L^p}{2} \left. \frac{d}{ds} \zeta_p^r
(s)
\right|_{s=0}, \nonumber \\
\zeta_p^r (s) &=&  \sum_{n=-\infty}^{\infty} \int_{-
\infty}^{+\infty} \frac{dk_1\cdots dk_p}{(2\pi)^p}   \left[
\frac{1}{\rho^2} \left(
k_1^2+ \cdots  +k_p^2 +
\frac{4\pi^2 n^2}{\beta^2t^2}\right)^2 \right. \nonumber \\
&+& \left. \kappa \left( \lambda_1
(k_1^2+ \cdots  +k_p^2  ) +
\frac{4\pi^2\lambda_0 n^2}{\beta^2t^2} \right)  \right]^{-s}\nonumber \\
&=&  \frac{V_p\rho^{2s}}{2(2\pi)^p}  \sum_{n=-\infty}^{\infty}
\int_0^{\infty} dk \, k^{p/2-1}(k+c_+)^{-s} (k+c_-)^{-s},
\end{eqnarray}
where $V_p$ is again the `volume' of the $p-1$-dimensional unit sphere
and the $c_{\pm}$ are again given by (\ref{cpm}).
A point about dimensions: the reader may verify (see also (38)) that
$[\kappa ] = M^{p+1}$ and $[\rho ]= M^{(1-p)/2}$.

 Proceeding as above,
namely considering the $n=0$ term separately, we obtain the
following generalization of the formula corresponding to the rigid
membrane:
\begin{eqnarray}
\zeta_p^r (s)&=& \frac{V_p\rho^{p-2s}(\kappa \lambda_1)^{p/2-
2s}}{2(2\pi)^p} \, \frac{ \Gamma (p/2-s) \Gamma (2s-p/2)}{\Gamma
(s)} +  \frac{V_p\rho^{2s}}{(2\pi)^p} \nonumber \\ &\times &\frac{ \Gamma
(2s-p/2) \Gamma (p/2)}{\Gamma (2s)}
\sum_{n=1}^\infty c_-^{p/2-2s} F(s,2s-p/2;2s;1-c_+/c_-).
\label{rpb0}
\end{eqnarray}
When computing its derivative at $s=0$ we must distinguish the two
cases: $p$ even and $p$ odd. For odd $p$ (which is the
simplest case):
\begin{eqnarray}
 \left. \frac{d}{ds} \zeta_p^r (s) \right|_{s=0}
 &=& \frac{V_p  \Gamma (p/2) \Gamma (-p/2)}{2(2\pi)^p} \left\{
(\kappa \rho^2 \lambda_1)^{p/2} + 4 \zeta (-p) \left(
\frac{2\pi}{\beta t} \right)^p \right.  \\ &+& \left. p \, \zeta
(2-p)\, \kappa \rho^2 \left[ \frac{p}{2} \lambda_1+ \left( 1-
\frac{p}{2} \right) \lambda_0 \right] \left( \frac{2\pi}{\beta t}
\right)^{p-2} \right\}, \ \ \ \ \
 p = \mbox{odd}, \ \ \  p\geq 3.       \nonumber
\end{eqnarray}
However, for even $p$ we must take into account that a divergence
in $\Gamma (s-p/2)$ can compensate the divergence of $\Gamma (s)$
in the denominator, when $s\rightarrow 0$. One should be careful
whith this fact, that renders the calculation more involved. The
result is
\begin{eqnarray}
 \left. \frac{d}{ds} \zeta_p^r (s) \right|_{s=0}
 &=& \frac{V_p  (-1)^{p/2}}{(2\pi)^p} \left\{ \frac{1}{p} (\kappa
\rho^2 \lambda_1)^{p/2} \left[ - \ln \left( \frac{\kappa \rho
\lambda_1}{\mu} \right) - \frac{\gamma + \psi (p/2)}{2} + S(p/2)
\right]  \right. \nonumber \\ &+& \frac{8}{p} \zeta' (-p)  \left(
\frac{2\pi}{\beta t} \right)^p -
 \frac{8}{p} \zeta (-p)  \left( \frac{2\pi}{\beta t} \right)^p
\ln \left( \frac{2\pi}{\beta t \sqrt{\rho}} \right) \nonumber \\ &-& 2 \zeta
(2-p) \kappa \rho^2 \left[ \frac{p}{2} \lambda_1+ \left( 1- \frac{p}{2}
\right) \lambda_0 \right] \left( \frac{2\pi}{\beta t} \right)^{p-2}
\ln \left( \frac{2\pi}{\beta t \sqrt{\rho}} \right) \nonumber \\ &+&  2 \zeta'
(2- p)  \kappa \rho^2 \left[ \frac{p}{2} \lambda_1+ \left( 1-
\frac{p}{2} \right) \lambda_0 \right] \left( \frac{2\pi}{\beta t}
\right)^{p-2} \nonumber \\
&+& \frac{4}{p} \zeta (-p) S(p/2)  \left( \frac{2\pi}{\beta
t} \right)^p +  \zeta (2-p) S(p/2-1) \kappa \rho^2 \lambda_1 \left(
\frac{2\pi}{\beta t} \right)^{p-2} \nonumber  \\ &+& \left. \left( \frac{p}{2}
-1 \right)
 \zeta (2-p) [S(p/2-2)-1]  \kappa \rho^2 (\lambda_1- \lambda_0)
\left( \frac{2\pi}{\beta t} \right)^{p-2} \right\}, \nonumber  \\  & &
 p = \mbox{even}, \ \ \  p\geq 4,
\end{eqnarray}
being $S(k)\equiv \sum_{l=1}^k l^{-1}$.
Finally, these expressions multiplied by $- \frac{1}{2} (d-p-1)L^p$
yield the one-loop contribution to the action of the rigid
 $p$-brane. In the case of the rigid membrane we get the rather
simpler result
\begin{equation}
 \left. \frac{d}{ds} \zeta_2^r (s) \right|_{s=0} =
 \frac{1}{4\pi } \kappa \rho^2 \lambda_1 \left[
\ln \left( \kappa \rho^2 \lambda_1 (\beta t)^2 \right) -1
\right] - \frac{ 8 \pi \zeta' (-2)}{(\beta t)^2}
 \frac{1}{4\pi } \kappa \rho^2 (\lambda_1 -\lambda_0).
\label{zrm}
\end{equation}
It is easy to check that the action corresponding to the rigid
membrane is just the
particular case $p=2$ of the rather non-trivial formula
corresponding to the rigid $p$-brane, for $p$ even. The full
power of zeta function regularization has been used in working out
such expression.

Again, it is remarkable that one can in fact obtain the {\it exact}
expressions corresponding to bosonic membranes and $p$-branes, both
for the ordinary and for the rigid case. The last one is really
intricate, specially for dealing with membranes and
 $p$-branes with $p$ even.

 The values of the extrema and of the Hagedorn
temperature will be derived after carefully taking  the high
temperature approximation consistently with the loop
expansion (that is, in the expansion (20) of the hypergeometric
function the first three terms must be kept).
\bigskip

\section{The minimal configuration and Hagedorn temperature for
bosonic membranes}

Again, the cases of the pure bosonic membrane and of the rigid
membrane will be considered.

\subsection{Pure bosonic membrane}

We start from the formula for the effective action of
the bosonic membrane to one-loop order:
\begin{equation}
S_{eff}^{(bm)} = \kappa L^2 \beta t \left[
(1+\sigma_0)^{1/2} (1+\sigma_1) - \left(
\frac{\lambda_0}{2} \sigma_0 +
\lambda_1\sigma_1\right) \right]-
\frac{2\pi (d-3)L^2 \zeta'(-2) \lambda_0}{\lambda_1 \beta^2t^2}.
\end{equation}
The conditions for extremum of this action are the following
\begin{eqnarray}
\lambda_0& =&(1+\sigma_0)^{-1/2} (1+\sigma_1), \nonumber \\
\lambda_1& =&(1+\sigma_0)^{1/2}, \nonumber \\
\lambda_1& =&- \frac{4\pi (d-3) \zeta' (-2)}{\kappa
\sigma_0 \beta^3t^3}, \\
\kappa \beta t \sigma_1 &=& \frac{2\pi (d-3) \zeta'(-2)
\lambda_0}{\lambda_1^2 \beta^2t^2}, \nonumber
\end{eqnarray}
Solving these equations for high temperature yields the
following behavior for the parameters
\begin{eqnarray}
\sigma_0 & \simeq & u^{2/3} (\beta t)^{-2} -\frac{1}{3}, \nonumber \\
\lambda_1 & \simeq & -u^{1/3} (\beta t)^{-1} -
\frac{1}{3} u^{-1/3} \beta t, \nonumber \\
\sigma_1 & \simeq & -\frac{1}{3}+
\frac{2}{9} u^{-2/3} (\beta t)^2, \\
\lambda_0 & \simeq & -\frac{2}{3} u^{-1/3}  \beta t, \nonumber
\end{eqnarray}
plus terms of higher order in $\beta$ and $\kappa$, being
\begin{equation}
u \equiv \frac{4\pi (d-3) \zeta'(-2)}{\kappa}.
\end{equation}
The asymptotic behavior of the winding soliton mass for
high temperature is easily found to be
\begin{equation}
M_1^{(bm)} \simeq \kappa^{2/3} \left[ 4\pi
(d-3)|\zeta'(-2)| \right]^{1/3}
+ \frac{1}{3}\, \kappa^{4/3} \left[ 4\pi (d-3)|\zeta'(-2)|
 \right]^{-1/3} (\beta t)^2.
\end{equation}
The same type of
 behavior as for the Nambu-Goto string shows up here:
the mass does not feel the increae in temperature and remains
constant, while corrections are damped by a second inverse
power of $T$. One could try to identify in this last
expression the Hagedorn temperature as the inverse value
of $\beta$ for which the soliton mass vanishes. However
this would yield a value proportional to the coupling
constant $\kappa$, which is considered to be small, so
that the whole approach would not make any sense. We shall
see that the situation in this respect is completely
different in the case of the rigid membrane. There,
the ideas of Polchinski et al. corresponding to the
rigid string can in fact be  implemented.
\medskip

\subsection{Rigid membrane}

The one loop action for the rigid membrane is readily obtained from
(\ref{zrm})
\begin{eqnarray}
S_1^{(rm)} &=& \frac{(d-3)L^2}{8\pi} \, \kappa \rho^2\lambda_1
\left[ 1 - \ln \left( \kappa \rho^{2} \lambda_1 (\beta t)^2\right)
\right] +   \frac{4\pi (d-3)L^2}{(\beta
t)^2} \, \zeta' (-2) \nonumber \\
 &+&  \frac{(d-3)L^2}{8\pi} \, \kappa \rho^2 (\lambda_1-\lambda_0).
\label{s1rm}
\end{eqnarray}
Here, terms up to $k=2$ in the expansion (20) of the hypergeometric
function of (\ref{rpb0}) have been taken
into account. It can be checked that we are thus absolutely consistent
 ---to first order in
the coupling constant $\kappa$--- with the high-temperature
 expansion
(the suppressed contributions are at least of order $L^2\kappa^{3/2}
\rho^3 \beta$). Notice, however, that all higher-order terms
would be easy to obtain from (\ref{rpb0}) and that a consistent
loop
expansion to any desired order can in fact be performed.

The conditions for extremum of $S^{(rm)} = S_0^{(m)}+
 S_1^{(rm)}$, eqs.
(\ref{tlm}) and (\ref{s1rm}), are obtained by taking the
derivatives with respect to the parameters  $\lambda_0$,
$\lambda_1$, $\sigma_0$ and  $\sigma_1$. The result is
\begin{eqnarray}
\sigma_0 & =& - \frac{d-3}{4\pi} \, \frac{\rho^2}{\beta t}, \nonumber  \\
  \sigma_1 &=&  \frac{(d-3)\rho^2}{8\pi \beta t} \left[
1 - \ln \left( \kappa \rho^{2} \lambda_1 (\beta t)^2\right)
\right], \nonumber \\
 \lambda_0 &=& (1+\sigma_1) (1+\sigma_0)^{-1/2}, \nonumber \\
 \lambda_1 &=&  (1+\sigma_0)^{1/2}.
\label{exrm}
\end{eqnarray}

Let us now analyze these equations.
We can easily identify here the transition that also takes place
for the rigid
string: for values of the temperature higher than the one
coming from the expression
\begin{equation}
\beta_{c}^{-1} = \frac{ 4\pi t}{(d-3)\rho^2},
\label{tcc}
\end{equation}
the values of the parameters, and hence of the action and of the
winding soliton mass squared, acquire an imaginary part. Guided by
the fact that in the rigid string case this temperature lies above
the Hagedorn temperature ---which, on the other hand, is supposed
to be high--- we conclude that in order that the whole scheme of
the string case can be translated to the membrane situation we must
demand that $\rho^2\mu$ be small ($\mu$ a typical mass scale).

It is remarkable that this necessary condition turns out to be
sufficient in order to get sensible results. In fact,
for the Hagedorn temperature, defined as the value for which the
winding soliton mass
\begin{equation}
M_1^{(rm)} \equiv \frac{S_{eff}^{(rm)}}{L^2}
\end{equation}
vanishes, we find
\begin{equation}
\beta_H^{-1} \simeq \frac{ 4\pi \sqrt{-\zeta'(-2)}\   t}{\rho
\sqrt{-\kappa \ln (\kappa \rho^6)} }.
\label{tcH}
\end{equation}

Before going on with the analysis, the following remark is
in order. The values
of the constants which determine the leading behavior of the
effective
action at high temperature, namely the derivative of the zeta
function at the point $-2$ (in general,  $-p$, respectively), have
been
calculated with unchallenged precision by one of the authors in
ref. [22] (in fact, asymptotic expansions for the derivatives to
any order of the zeta function, which provide extremely accurate
numerical results with just a few terms of the series, have been
given there). In particular
\begin{equation}
\zeta' (-1) = -0.16542115, \ \ \ \ \  \zeta' (-2) = -0.03049103.
\end{equation}

We are now in the position to analyze the evolution of the
winding-soliton mass for the rigid string at one loop order:
\begin{equation}
M_1^{(rm)} \simeq \frac{4\pi (d-3) \zeta' (-2)}{(\beta t)^2}.
\end{equation}
Let us recall that we are assuming $\rho^2$  small. For the sake
of precision, let us take it to be of order $\kappa$.
That is, $\kappa \mu^{-3} \sim \rho^2\mu$, and for the adimensional
quantity $\kappa\rho^6 <<1$. When the temperature is varied from slow to
higher values,
the winding-soliton mass starts from its classical value near $\kappa
\beta t$ and ends at minus infinity. Similarly to what happens to the
rigid
string, in the meantime, as the temperature increases two different
critical values can be identified. The first one occurs when the
winding-soliton mass goes through zero. This corresponds to the
Hagedorn transition and is obtained at the value
\begin{equation}
 \beta_H^{-1}  \simeq \frac{ t}{\rho^2 \mu^2 \kappa \sqrt{-\ln
(\rho^2 \mu) }}. \end{equation}
The second privileged value of the temperature occurs at $T_c$ as
given by expression (\ref{tcc}). This value lies above the
Hagedorn temperature and signalizes the appearance of an imaginary
part in the soliton mass squared, as happened with the rigid string.

 We can readily check that everything is correct in this regime: (i) the
complex transition takes place for {\it high temperature}; (ii) the
Hagedorn transition takes place {\it before} the complex
transition; (iii) in the Hagedorn transition the behavior of the
constants is completely {\it consistent} with the high-temperature
expansion, in fact, $\sigma_0$ and $\lambda_1$ are of order 1,
while  $\sigma_1$ and $\lambda_0$ are logarithmically divergent.
More precisely, assuming typical values for the constants,  such as
$\rho \mu =10^{-1}$
and $\kappa \mu^{-3} \simeq \rho^2 \mu =10^{-2}$, the values of the
temperature at the two transition points are
\begin{equation}
\beta_H^{-1} \simeq 10 \pi, \ \ \ \
\beta_c^{-1} \simeq \frac{4\pi \cdot 10^2}{d-3}.
\end{equation}

 Substituting into (\ref{s1rm}) the values of the parameters
at the extremum, we can check that we (consistently) obtain again (61),
which is apparently different from the large-$N$ QCD
result (although the sign is coincident).

\bigskip

\section{Discussion and conclusions}

The extension of the method of zeta function regularization
developed in [12] has been used in the calculation of the partition
functions for strings and membranes.
These results are essential in order to decide,
once and forever, if QCD can possibly be a certain limit of some
string
(or membrane, or $p$-brane) theory, through the analysis of the high
temperature behavior of the corresponding partition functions.

We have developed a method that allows the explicit
calculation of the partition functions of all these theories of
extended
objects at any value of the temperature and to any desired order of
approximation, always consistently with the loop expansion.
Although the
procedure involves  rather elaborate mathematical
tools (mainly the so-called zeta function regularization
techniques), the final results are
expressed in terms of simple series, whose first few terms give
normally
the desired result. We can also take advantage of the
high-precision calculations of the derivatives of the zeta
functions that are available in the literature.

The two transition which were shown in
ref. [5] to occur in the case of the rigid string, for increasing
temperature, have been also clearly characterized here in the case
of the rigid membrane. In particular, assuming a reasonable
behavior of the parameters of the model, we have been able to
define
the Hagedorn transition in general, for the rigid membrane
and $p$-brane. And this has always been a very elusive concept.

The high-temperature limits of the free energy per unit area for
the bosonic membrane and for the rigid membrane have been found not
to agree with the high temperature behavior
of large-$N$ QCD. In fact, in the case of the bosonic membrane
it remains constant, suffering from the same problem as the
Nambu-Goto
string. On the opposite side, in the case of the rigid membrane it
grows too quickly with $\beta^{-1}$ (namely,
as $\beta^{-2}$, contrary to what happens with the Nambu-Goto
string and bosonic membrane, which remain constant with $T$). Only
the rigid string seems to yield the desired behavior $\beta^{-1}$.
With our method,   a systematic analysis
of all the distinct cases corresponding to the different kinds of
membranes and $p$-branes has been rendered possible.

\vspace{5mm}
\noindent{\large \bf Acknowledgments}

Financial support from DGICYT (Spain), research project
PB90-0022, from the Generalitat de Catalunya and from the Alexander
von
Humboldt Foundation (Germany) is gratefully acknowledged.
The paper was finished during a short visit of EE at Trento
University. He thanks A.A. Bytsenko, G. Cognola, K. Kirsten,
L. Vanzo and S. Zerbini for discussions.
SDO wishes to thank JSPS (Japan) for partial financial
support of this work.
 \bigskip


\newpage



\begin{thebibliography}{99}

\bibitem{} G. 't Hooft, Nucl. Phys. {\bf B72} (1974) 461.
\bibitem{} J. Polchinski, preprint UTTG-16-92 (1992).
\bibitem{} D. Gross, preprint PUPT-1355 (1992).
 \bibitem{}   J. Polchinski,  Phys. Rev. Lett. {\bf 68} (1992)
1267.
 \bibitem{}   J. Polchinski and Z. Yang,  Phys. Rev.
{\bf D46} (1992) 3667.
\bibitem{} A.M. Polyakov, Nucl. Phys. {\bf B268} (1986) 406.
\bibitem{} H. Kleinert, Phys. Lett. {\bf B174} (1986) 335.
\bibitem{} E. Braaten, R. Pisarski and S. Tze, Phys. Rev. Lett.
{\bf 58} (1987) 93;
  T.L. Curtright, G.I. Ghandhour and C.K. Zachos, Phys. Rev.
{\bf D34} (1986) 3811;
  F. Alonso and D. Espriu, Nucl. Phys. {\bf B283} (1987) 393;
 H. Kleinert, Phys. Rev. Lett. {\bf 58} (1987) 1915;
 P. Olesen and S.K. Yang, Nucl. Phys. {\bf B283} (1987) 74;
R. Pisarski, Phys. Rev. Lett. {\bf 58} (1987) 1300;
E. Braaten and C.K. Zachos, Phys. Rev. {\bf D35} (1987) 1512.
\bibitem{} G. German and H. Kleinert, Phys. Lett. {\bf B225}
(1989) 107.
\bibitem{}
 S.W. Hawking, Commun. Math. Phys. {\bf 55} (1977) 133;
 J.S. Dowker and R. Critchley,  Phys. Rev. {\bf D13} (1976)
3224;
L.S. Brown and G.J. MacLay,  Phys. Rev. {\bf 184} (1969) 1272.
\bibitem{}
 E. Elizalde,  J. Math. Phys. {\bf 31} (1990) 170;
J. Phys. {\bf A22} (1989) 931.
\bibitem{}
E. Elizalde, S. Leseduarte and S. Zerbini, {\it
Mellin transform techniques for zeta-function resummations},
Barcelona Univ. preprint (1993).

\bibitem{}
 H.A. Weldon,  Nucl. Phys. {\bf B270 [FS 16]} (1986) 79;
A. Actor,  Fortschr. Phys. {\bf 35} (1987) 793;
A. Actor, J. Phys. {\bf A24} (1991) 3741;
S.K. Blau, M. Wisser and A. Wipf, Nucl. Phys. {\bf B310}  (1988)
163;
P. Chang and J.S. Dowker, Manchester preprint (1992).
\bibitem{} S.D. Odintsov, Riv. Nuovo Cim. {\bf 15} (1992) 1.
\bibitem{}
P.A.M. Dirac, Proc. R. Soc. {\bf A268} (1962) 57;
P.S. Howe and R.W. Tucker, J. Phys. {\bf A10} (1977) L155;
A. Sugamoto, Nucl. Phys. {\bf B215} (1983) 381.
\bibitem{} E.G. Floratos and G.K. Leontaris, Phys. Lett. {\bf B223}
(1989) 137.
\bibitem{} S.D. Odintsov and D.L. Wiltshire, Class. Quant. Grav.
{\bf 7} (1990) 1499;
A.A. Bytsenko and S.D. Odintsov, Class. Quant. Grav. {\bf 9}
 (1992) 391;
A.A. Bytsenko and S. Zerbini, Trento Univ. preprint 1992;
A.A. Bytsenko, K. Kirsten and S. Zerbini, Phys. Lett.
{\bf B}, to appear.
\bibitem{} M.J. Duff, T. Inami, C.N. Pope, E. Sezgin and K.S.
Stelle,
Nucl. Phys. {\bf B227} (1988) 515;
K. Fujikawa and J. Kubo, Phys. Lett. {\bf B199} (1987) 75.
\bibitem{} E. Bergshoeff, E. Sezgin and P.K. Townsend, Ann. Phys.
(NY)
 {\bf 185} (1988) 330.
\bibitem{} S.D. Odintsov, Phys. Lett. {\bf B247} (1990) 21;
E. Elizalde and S.D. Odintsov, Phys. Rev. D, to appear.
\bibitem{} A.A. Bytsenko and S.D. Odintsov, Fortschr. Phys.,
to appear.
\bibitem{} E. Elizalde, {\it A simple recurrence for the higher
derivatives of the Hurwitz zeta function}, J. Math. Phys., to
appear.

\bibitem{} M.J. Duff, Class. Quant. Grav. {\bf 6} (1989) 2577.

\bibitem{} M.J. Duff, P.S. Howe, T. Inami and K. Stelle, Phys.
Lett.  {\bf B191}, 70 (1987);
A. Ach\'{u}carro, P. Kapusta and K.S. Stelle,  Phys.
Lett.
{\bf B232}, 302 (1989); U. Lindstr\"{o}m, Stockholm preprint, 1990.


\end{thebibliography}
\end{document}